\documentclass[preprint,10pt]{elsarticle}
\usepackage{threeparttable}
\usepackage{microtype}
\usepackage{multirow}
\usepackage{tablefootnote}
\usepackage{microtype}
\usepackage{booktabs}
\usepackage{xcolor}
\usepackage{float}
\usepackage{gensymb}
\usepackage{amssymb}
\usepackage{amsmath}
\usepackage{siunitx}
\usepackage[version=4]{mhchem}
\usepackage[english]{babel}
\renewcommand{\textcolor}[1]{#1}

\begin{document}

\begin{frontmatter}
\title{Weyl Semimetallic Phase in High Pressure \ce{CrSb2} and Structural Compression Studies of its High Pressure Polymorphs}

\author[1]{Carl Jonas Linnemann}
\affiliation[1]{organization={Department of Chemistry and iNANO, Aarhus University},addressline={Langelandsgade 140},postcode = {8000},city={Aarhus C}, country = {Denmark}}
\author[1,2]{Emma Ehrenreich-Petersen}
\affiliation[2]{organization = {Deutsches Elektronen-Synchrotron DESY}, addressline = {Notkestr. 85}, postcode = {22607}, city = {Hamburg}, country = {Germany}}
\author[3]{Davide Ceresoli}
\affiliation[3]{organization = {Consiglio Nazionale delle Ricerche - Istituto di Scienze e Tecnologie Chimiche ``G. Natta''}, addressline = {via Golgi 19}, postcode = {20133}, city = {Milano}, country = {Italy}}
\author[2]{Timofey Fedotenko}

\author[4]{Innokenty Kantor}
\affiliation[4]{organization = {MAX IV Laboratory, Lund University},addressline = {Fotongatan 2},city = {Lund}, postcode = {225 94},country = {Sweden}}

\author[1,4]{Mads Ry Vogel Jørgensen}

\author[1]{Martin Bremholm \corref{cor1}}
\ead{bremholm@chem.au.dk}
\cortext[cor1]{Corresponding author}
\begin{abstract}
In this study, high pressure synchrotron powder X-ray diffraction is used to investigate the compression of two high pressure polymorphs of \ce{CrSb2}. The first is the \ce{CuAl2}-type polymorph with an eight-fold coordinated Cr, which can be quenched to ambient conditions from high-pressure high-temperature conditions. The second is the recently discovered \ce{MoP2}-type polymorph, which is induced by compression at room temperature, with a seven-fold coordinated Cr. Here, the assigned structure is unambiguously confirmed by solving it from single-crystal X-ray diffraction. Furthermore, the electrical properties of the \ce{MoP2}-type polymorph were investigated theoretically and the resistance calculations under pressure were accompanied by resistance measurements under high pressure on a single crystal of \ce{CrSb2}. The calculated electronic band structure for the \ce{MoP2}-type phase is discussed and we show that the polymorph is semimetallic and possesses type-I Weyl points. No further phase transitions were observed for the \ce{CuAl2}-type structure up to 50 GPa and 40 GPa for the \ce{MoP2}-type structure. Even though the \ce{CuAl2}-phase has the highest coordination number of Cr, it was found to be less compressible than the \ce{MoP2}-phase having a seven-fold coordinated Cr, which was attributed to the longer Cr-Sb distance in the \ce{CuAl2}-type phase. The discovery of a type-I Weyl semimetallic phase \textcolor{red}{in \ce{CrSb2}} opens up for discovering other Weyl semimetals in the transition metal di-pnictides under high pressure. 
\end{abstract}

\begin{keyword}
Crystal binding and equation of state\sep Electronic band structure\sep Phase transitions
 \sep High pressure\sep X-ray diffraction
\end{keyword}

\end{frontmatter}

\section{Introduction}
\begin{figure}[h]
\centering
    \includegraphics[width=\linewidth]{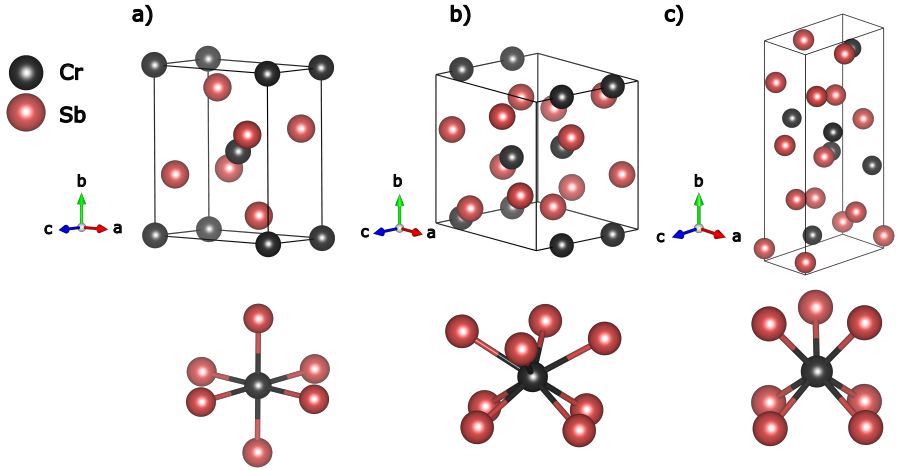}
\caption{Crystal structure and Cr coordination of the different high-pressure polymorphs of \ce{CrSb2}: \textbf{a)} Marcasite, \textbf{b)} \ce{CuAl2}-type (HP-I) and \textbf{c)} \ce{MoP2}-type (HP-II).}
    \label{fig:structures}
\end{figure}

Many \textcolor{red}{compounds} exist in different polymorphs with different properties, such as \ce{WP2}, which exists in the low temperature modification $\alpha$-\ce{WP2} \cite{alphawP2} and the high temperature modification $\beta$-\ce{WP2} crystallizing in the $Cmc2_1$ space group \cite{MoP2WP2}. Both polymorphs are semimetals where $\beta$-\ce{WP2} is a type-II Weyl semimetal and $\alpha$-\ce{WP2} is a trivial semimetal \cite{MoP2WP2_weyl,alphawp2MR}. Pressure can also induce phase transitions such as in \ce{CrSb2} where, recently, a new high pressure polymorph was discovered independently by Li et al. as well as Ehrenreich-Petersen et al., extending the number of known polymorphs of \ce{CrSb2} to three \cite{KinesiskHP-II,<EmmaMoP2>}.
At ambient pressure \ce{CrSb2} crystallizes in the marcasite structure (Fig. \ref{fig:structures} a)) and has an antiferromagnetic (AFM) ordering below $T_N\approx\SI{273}{K}$, with a 1x2x2 supercell with the magnetic moments perpendicular to the (101) plane \cite{<CrSb2_neutron>}. The marcasite phase is a $n$-type semiconductor with a narrow bandgap of \SI{0.1}{eV} \textcolor{red}{and has been shown to possess large anisotropic magnetoresistance in its surface states \cite{<Physb2012>,CrSb2_surf_MR}}.
The first high-pressure polymorph of \ce{CrSb2} (herein named HP-I) was discovered by Takizawa et al. by high pressure synthesis  above \SI{5.5}{GPa} and \SI{600}{\celsius} \cite{<Takizawa_1>}. This polymorph crystallizes in a \ce{CuAl2}-type structure (Fig. \ref{fig:structures} b)), \textcolor{red}{in which \ce{TiSb2} and \ce{VSb2} also crystallize \cite{TiSb2_VSb2}. Here, \ce{TiSb2} has been shown to be a three-dimensional Dirac-semimetal \cite{TiSb2_dirac}}. \textcolor{red}{The HP-I phase of \ce{CrSb2}} exhibits metallic behaviour and is an itinerant electron ferromagnet with a Curie temperature of approximately \SI{160}{K} \cite{<HPCrSb2Takizawa>}. At approximately \SI{90}{K} it has an AFM spin density wave transition, which can be suppressed under the application of magnetic fields. It has been shown that at pressures above \SI{0.5}{GPa} the two states merge together to a single AFM state, where the AFM order is suppressed to lower temperatures by application of pressure and vanishes completely at $P_c\approx\SI{9}{GPa}$, indicating an AFM quantum critical point \cite{<Jiao_HP_I>}.
The newly discovered high pressure polymorph (herein named HP-II) has been shown to appear at room temperature and pressures above \SI{10}{GPa}. It is meta-stable and cannot be recovered to ambient pressure \cite{KinesiskHP-II,<EmmaMoP2>}. Furthermore laser-heating experiments under high pressure have shown that the phase transforms into the high-pressure \ce{CuAl2}-type phase, indicating a high activation barrier for the phase transition to the HP-I phase and that it is more stable than the HP-II phase \cite{<EmmaMoP2>}. Li et al. showed that the polymorph exhibits \textit{n-p} conduction switching as well as that the high pressure phase exhibits metallic conduction \cite{KinesiskHP-II}. There have been disagreements regarding the structure, where Li et al. assigned it to the arsenopyrite structure-type, while Ehrenreich-Petersen et al. assigned it to have a \ce{MoP2}-type structure using a combination of DFT calculations and powder diffraction \cite{KinesiskHP-II,<EmmaMoP2>}. The \ce{MoP2}-type structure has the orthorombic $Cmc2_1$ space-group, where Mo is coordinated to seven P atoms. This structure is furthermore isostructural with $\beta$-\ce{WP2}, and both \ce{MoP2} and $\beta$-\ce{WP2} have been studied immensely as they are robust type-II Weyl-semimetals \cite{MoP2WP2,MoP2WP2_weyl}. Both materials have been shown to possess extremely high magnetoresistance and high conductivity\cite{Mr-wP2_MoP2}. \textcolor{red}{Furthermore, high harmonic generation up to the 10\textsuperscript{th} order has been shown for $\beta$-\ce{WP2} \cite{WP2_HHG}} and due to their high magnetoresistance both materials are potential candidates for high-gain high-frequency amplifiers with low power dissipation, which could replace High-electron-mobility transistors in quantum computers \cite{Amplifier}. \textcolor{red}{For both compounds no structural phase transitions under the application of pressure have been shown, however $\beta$-\ce{WP2} has been shown to possess a pressure induced Lifshitz transition \cite{Alpha_MoP2,WP2_high_pressure}. Both \ce{MoP2} and \ce{WP2} have also shown to crystallize in the monoclinic \ce{OsGe2}-type structure. For \ce{WP2} it is a low temperature modification, while it for \ce{MoP2} has been shown to be a recoverable high-pressure high-temperature polymorph \cite{MoP2WP2,Alpha_MoP2}. }

\textcolor{red}{In this study the high pressure structural properties of both high pressure polymorphs of \ce{CrSb2} are investigated experimentally through powder diffraction at high pressures as well as theoretically through density functional theory (DFT) calculations. Furthermore, the \ce{MoP2}-type structure is shown to be the correct structure for the HP-II polymorph of \ce{CrSb2} through single crystal X-ray diffraction at 12 GPa. The electronic properties of this polymorph are furthermore investigated through resistance measurements under high pressure at ambient temperature accompanied by theoretical calculations regarding the resistivity and Hall coefficients. The band structure is calculated at 12 GPa and the topology of the band crossings is determined to examine if the HP-II phase possesses Weyl points.} The verification of the \ce{MoP2}-type structure, and furthermore the evidence of Weyl-points opens up for discovering other high pressure transition-metal di-pnictides, which potentially host Weyl points.

\section{Materials and methods}
\subsection{Powder synthesis of \ce{CrSb2}}
Stoichiometric amounts of \ce{Cr} (99.95\%) and \ce{Sb} (99.95\%) were ground in a mortar, pressed to a pellet and sealed in an evacuated quartz tube, which was heated at a rate of $\SI{50}{\degree C}$/min to  \SI{600}{\degree C}, kept at that temperature for \SI{48}{h} and then quenched in air.

\subsection{Single crystal growth of \ce{CrSb2}}
Single crystals of \ce{CrSb2} were grown using a flux method adapted from Ref. \cite{<Physb2012>}. \ce{Cr} (99.95\%) and \ce{Sb} (99.99\%) with a molar ratio of 6:94 were ground in a mortar and sealed in an evacuated quartz tube. The tube was heated to \SI{1000}{\degree C} at $\SI{50}{\degree C}$/min, kept at that temperature for \SI{36}{h} followed by cooling to \SI{640}{\degree C} at $\SI{2}{\degree C}$/h and then cooled to room temperature. The crystals were isolated from the Sb-flux through centrifugation. The phase purity of the crystals was verified by determining the unit cell of a single crystal using a STOE Stadivari single crystal diffractometer with Mo K-$\alpha$ radiation (Table S1).

\subsection{High pressure synthesis of \ce{CuAl2}-type \ce{CrSb2}}
A pellet of marcasite \ce{CrSb2} powder was loaded in a large volume press in a 14/8 COMPRES octahedral assembly \cite{COMPRES}, pressurized to \SI{7}{GPa} at \SI{1}{{GPa}/{h}} and heated to \SI{900}{\degree C} for \SI{1}{h} followed by quenching. The phase purity of the sample was verified using powder X-ray diffraction using a Rigaku Smartlab diffractometer with Co K-$\alpha_{1,2}$ radiation (Fig. S1).

\subsection{High pressure diffraction measurements}
Single-crystal X-ray diffraction (SC-XRD) measurements under high pressure were conducted at the Extreme Conditions Beamline (ECB, P02.2)\cite{P022}, PETRA III, DESY ($\lambda$=0.2899 Å, beam size $\approx 2\times2.7$ \si{\micro m^2}, Perkin Elmer XRD 1621 detector). Three single crystals of \ce{CrSb2} were loaded in a plate-diamond anvil cell (DAC) (Almax EasyLab) with \SI{600}{\micro m} culet diamonds. Additionally, a small ruby sphere was loaded that served as the pressure calibrant during the experiment by the ruby fluorescence method \cite{<Syassen>}. A 16:3:1 mixture of methanol-ethanol-water was used as the pressure transmitting medium (PTM). 
A stainless-steel gasket pre-indented to \SI{72}{\micro m} micron with a \SI{300}{\micro m} hole drilled by electric discharge machining (EDM), served as the sample chamber. 
The pressure was increased to 12 GPa, i.e. above the structural phase transition of \ce{CrSb2} and the DAC was heated to \SI{100}{\celsius} for one hour to relieve strain. The pressure was allowed to equilibrate for several hours before the measurement and was measured right before and after the diffraction experiment with no observed pressure deviation. As most of the single crystals had split into separate crystal domains during the phase transition a suitable spot for diffraction was chosen and diffractograms were obtained with a rotation from $-37\degree$ to $37\degree$ with $0.5\degree$ steps with an acquisition time of 0.5 s and a \SI{50}{\micro m} Pt absorber inserted in the beam.
Reduction and integration of the data was performed using CrysAlisPro \cite{cryspro} and the structure was solved using SHELXT in Olex2 and refined using SHELXL \cite{Olex2,Shelxl,Shelxt}. During integration and reduction of the data, frames with a $R_{\mathrm{int}}$ above 0.3 were rejected. 

Membrane driven Mao-Bell type DACs with culet sizes of \SI{300}{\micro m} (For the HP-I phase) or \SI{200}{\micro m} (For the HP-II phase) were used for the high pressure powder X-ray diffraction measurements (HP-PXRD). \ce{Re} gaskets indented from approx. \SI{250}{\micro m} to approx. \SI{40}{\micro m} were used with a \SI{120}{\micro m} hole drilled by EDM for the HP-I phase and a laser-drilled \SI{100}{\micro m} hole for the HP-II phase. Silicone oil was used as the PTM for the HP-I phase and for the HP-II phase \ce{Ne} was gas-loaded as the PTM using the gas-loading system at Sector 13, APS \cite{<GasLoad>}.
To obtain small grain sizes, the samples were suspended in ethanol and the supernatant was removed and dried. Thin sheets of samples were loaded together with thin sheets of \ce{Cu} (for the HP-I phase) or \ce{Au} (for the HP-II phase) for pressure determination \cite{<CuEoS>,AuEoS}. A small ruby sphere was also added, which served to determine the pressure during the loading and closing of the cell.
For the HP-I phase the HP-PXRD measurements were conducted at the DanMAX beamline at the MAX IV synchrotron ($\lambda$=0.3542 Å, beam size $\approx 7.4\times7.2$ \si{\micro m^2}, Dectris Pilatus3X 2M CdTe detector). The diffractograms for the HP-II phase have been shown, but not refined, in Ref. \cite{<EmmaMoP2>}, and the HP-PXRD measurements for the HP-II phase of \ce{CrSb2} were conducted at the Advanced Photon Source at the GSECARS beamline 13-ID-D ($\lambda$=0.3344 Å, beam size $\approx4\times3\,\si{\micro m^2}$, Dectris Pilatus CdTe 1M detector). 
All 2D powder diffractograms were integrated using Dioptas \cite{<Dioptas>}. Le Bail refinements of the unit cell parameters were performed using the FullProf suite \cite{Fullprof} and the unit cell parameters were fitted to the third order Birch-Murnaghan equation of state (EoS) using the EoSFit7 GUI \cite{EoSfitGUI}. 
Crystal structure models were drawn using the VESTA software \cite{<vesta>}.

\subsection{Resistance measurement of \ce{CrSb2} under pressure}
The resistance measurements were performed using an iDiamond (Almax Easylab) containing eight tungsten microprobes partially embedded in an epitaxially grown diamond \cite{<First_designeranvil>}. The culet size of the anvil was \SI{250}{\micro m} with bevels.
A BX-90 DAC was used with a stainless-steel gasket indented from \SI{250}{\micro m} to \SI{50}{\micro m} using two standard diamond anvils \textcolor{red}{in order not to damage the electrodes on the iDiamond}. A hole with a diameter of \SI{120}{\micro m} was drilled using EDM. Steatite was used as the pressure transmitting medium.
A suitable single crystal of \ce{CrSb2} was mounted on the designer anvil together with a ruby for pressure determination \cite{<Syassen>} and the resistance was measured in a two-wire \textcolor{red}{direct-current} geometry with a current of 1 mA using a Keithley 2450 sourcemeter.

\subsection{DFT modeling}
DFT calculations were performed with the plane wave pseudopotential code Quantum Espresso~\cite{<quantespresso2009>,<quantespresso2017>} with ultrasoft pseudopotentials and the PBEsol exchange-correlation functional~\cite{<functionale>}. The plane wave and density cutoffs were 45 and 450~Ry, respectively. In all calculations we sampled the reciprocal space with a nearly constant density of k-points of $\simeq$0.2\,\AA$^{-1}$. We used a Gaussian smearing of 0.005~Ry for the electronic temperature. Within our DFT setup, the band gap of the AMFe marcasite structure is 0.06~eV, which slightly underestimates the experimental value of 0.07~eV~\cite{Kuhn2013}.
A Hubbard U correction of 2.3~eV as in Ref.~\cite{Kuhn2013} results in a worse agreement with the experimental AFMe lattice parameters, and, contrary to Ref.~\cite{Kuhn2013}, the AFMe marcasite structure becomes metallic, with Sb-5$p$ orbital bands crossing the Fermi level.
As a function of pressure, variable-cell geometry optimizations of \ce{CrSb2} were performed in three magnetic ground states: non-magnetic (NM), ferromagnetic (FM), and antiferromagnetic (AFM). For the marcasite structure we used the AFMe ordering reported in Ref.~\cite{Kuhn2013}. For the HP-II structure we generated all AFM orderings compatible with a 2$\times$2$\times$2 supercell and selected the lowest energy one (which we named AFM1, Fig.~S2).
 
The electrical conductivity was calculated with the BoltzTraP2 code~\cite{boltztrap2} at 300~K in the Constant Relaxation Time Approximation (CRTA), with a 20$\times$20$\times$20 k-point mesh. The search for Weyl points requires the evaluation of the DFT Hamiltonian on an extremely dense k-point grid. For this purpose, we used PAOFLOW2~\cite{PAOFLOW2} to construct a tight-binding Hamiltonian from the self-consistent plane wave Hamiltonian. The BFGS algorithm is used to locate band crossings in reciprocal space. Finally, the chirality (i.e. the $\mathbb{Z}_2$ invariant) of each band crossing was calculated using Z2PACK~\cite{Z2PACK}.

\section{Results and discussion}
\subsection{The crystal structure of the HP-II phase obtained from single-crystal X-ray diffraction}
\begin{table}[h!]
    \centering
    \caption{Results from the single-crystal diffraction experiment compared to the data from DFT and powder diffraction from ref. \cite{<EmmaMoP2>}}
    \begin{threeparttable}
    \begin{tabular}{lll}
    \toprule
    &Single crystal diffraction & Ref. \cite{<EmmaMoP2>}\tnote{a} \\
    \midrule
        Pressure (GPa)&11.9(2)&12.0(7)\\
    Space group&$Cmc2_1$&$Cmc2_1$\\
    $a$ (Å)&3.0752(2)&3.06723(6)\\
    $b$ (Å)&  12.4476(19)&12.4495(5)\\
    $c$ (Å)&5.8741(2)&5.85682(9)\\
    $\alpha,\beta,\gamma$ ($\degree$)&90&90\\
    $V$ (Å$^3$)& 224.85(4)&223.645(11)\\
    \multicolumn{2}{l}{\text{\underline{Atomic coordinates}}}\\
    Cr&(0, 0.3942(5), 0.1490(6))&(0, 0.39431, 0.15191)\\
    Sb1&(0, 0.20702(17), 0.3545(2))&(0, 0.20806, 0.36102)\\
    Sb2&(0, 0.05627(18), 0)&(0, 0.05501, 0)\\
    \multicolumn{2}{l}{\text{\underline{Refinement statistics}}}\\
    Completeness&55.9\% \\
    $I/\sigma(I)$&95.7\\
    $R_1$(\%)&5.82\\
    $wR_2$ (\%)&16.10\\
    $R_{\mathrm{int}}$ (\%)&0.62\\
    GooF&1.150\\
    $\lambda$ (Å)& 0.2899\\
    \bottomrule
    \end{tabular}
    \begin{tablenotes}
        \item[a]{The unit cell was standardized to match the setting from the single-crystal diffraction experiment.} 
    \end{tablenotes}
    \end{threeparttable}
    \label{tab:scxrd}
\end{table}
The results from the SC-XRD refinement are shown in Table \ref{tab:scxrd} together with the results from the DFT-calculations and powder diffraction from Ehrenreich-Petersen et al. \cite{<EmmaMoP2>}. Here, the structural solution from the single crystals of \ce{CrSb2} at 12 GPa shows that the HP-II structure is of the \ce{MoP2}-type in agreement with our previous paper \cite{<EmmaMoP2>}. 

\subsection{Electrical properties of the HP-II phase}
\begin{figure}[h]
    \includegraphics[width=140 mm]{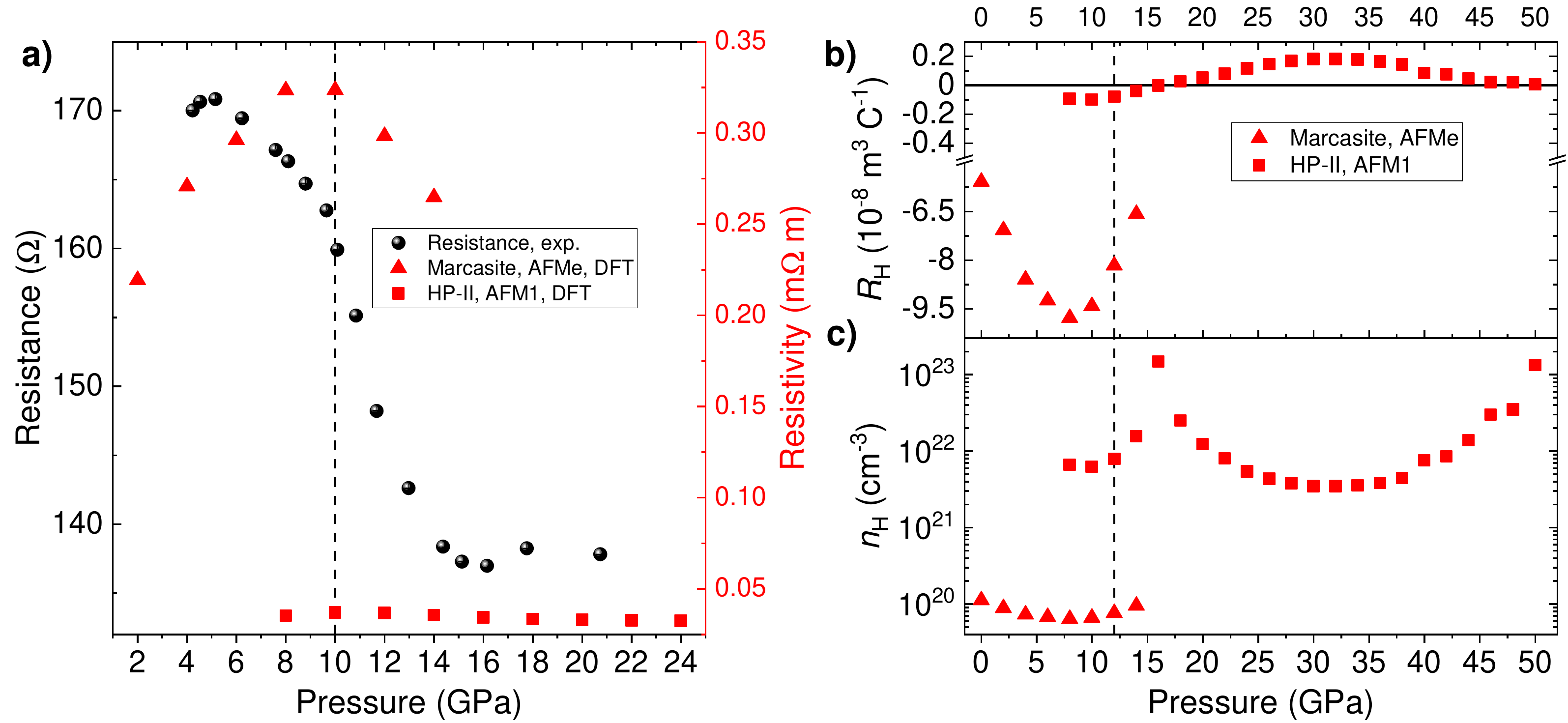}
\caption{a) Resistance of \ce{CrSb2} under high pressure: The experimentally measured resistance of \ce{CrSb2} under the phase transition to the \ce{MoP2}-structure is shown together with the calculated resistivity for the marcasite phase and the \ce{MoP2}-type phase. The dashed line marks the onset of the phase transition from the marcasite to the HP-II phase \cite{<EmmaMoP2>}. b) Calculated Hall coefficient. The dashed line marks the phase transition pressure from DFT \cite{<EmmaMoP2>} and the line at $R_\mathrm{H}=0$ serves as a guide to the eye. c) The carrier concentration calculated from the Hall coefficient.}
\label{fig:resi} 
\end{figure}

The electrical resistance of the HP-II phase was measured for pressures up to \SI{21}{GPa} and is shown in Fig. \ref{fig:resi}a) together with the calculated resistivity. Between 5 and 10 GPa a small linear decrease of the resistance is observed, which was also observed by Li et al. indicating a decrease of the band-gap of the marcasite phase with pressure, which also has been shown for some pressure intervals in \ce{FeSb2} \cite{<FeSb2Eg>,<FeSb2egexp>}. After the onset of the phase transition at 10~GPa \cite{<EmmaMoP2>}, a sharp decrease of the resistance is observed, indicating a metallization of the system. \textcolor{red}{The residual resistance of \SI{130}{\ohm} after the phase transition is high for a metal, which is attributed to the resistance in the electrodes, which for a similar anvil has been reported to be \SI{110}{\ohm} \cite{<First_designeranvil>}.}

\begin{figure*}[h]
    \centering
      \includegraphics[width=\textwidth]{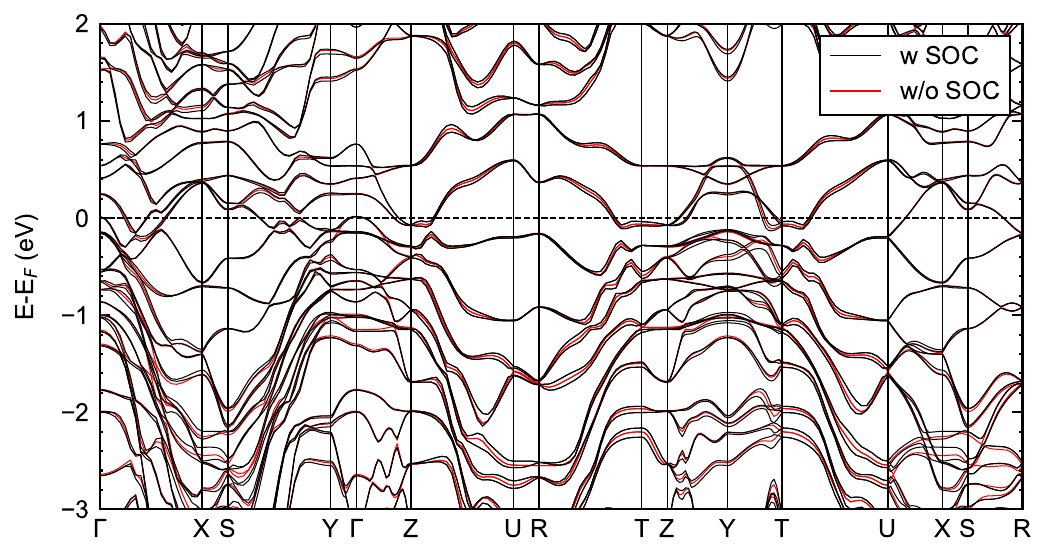} 
    \caption{Band structure of the HP-II AFM1 structure at 12 GPa around the Fermi level, with and without Spin Orbit Coupling (SOC).}\label{fig:bands}
\end{figure*}
The evolution of the electronic density of states as a function of pressure is shown in Fig.~S3 for the marcasite- and the HP-II polymorph. As reported in Ref.~\cite{Kuhn2013}, in the marcasite structure the magnetic order has a very strong impact on the electronic structure and the experimental AFMe order opens a $\sim$0.2~eV gap between the Cr-projected valence and conduction states. GGA calculations (as in the present case) predict a vanishing band gap, with a low density of states and low electron velocity. On the contrary GGA+U calculations~\cite{Kuhn2013} produce a narrow band of 0.35~eV, which overestimates the experimental band gap of $\sim$0.07~eV. Upon increasing pressure, the density of states reveals that a pseudogap is generated by the splitting of $3d$ Cr orbitals. In the marcasite structure, the band edges near the Fermi level have a large Cr-$3d$ character, while the Sb atomic orbitals contribute largely to the electronic properties below $-$2~eV and above $+$3~eV from the Fermi level (not shown in the figure). Overall, this results in a semi-metallic or semi-conducting behaviour. The situation is strikingly different for the HP-II polymorph. First, the system is metallic with a large density of states at the Fermi level, and the magnetic order has a modest impact on the electronic structure. The density of states of the HP-II structure at $E_{\mathrm
F}$ decreases as a function of pressure. The results thus confirm that the marcasite to HP-II transition is accompanied by a sudden metallization of the system. This leads to the lower resistivity of the HP-II phase compared to the marcasite phase. Furthermore, the drop of the resistance between 5 and 10 GPa for the marcasite phase is also reproduced by the calculations, however with a pressure offset of 4 GPa.
The calculated Hall-coefficient (Fig. \ref{fig:resi} b)) is in accordance with the measurements by Li et al. \cite{KinesiskHP-II} with a change to a positive Hall coefficient after the phase transition. The carrier concentration obtained from the Hall coefficient (Fig. \ref{fig:resi}  c)) is on the same order of magnitude as in Li et al., however our calculated maximum is 4 GPa lower than the observed maximum by Li et al.\cite{KinesiskHP-II} which is attributed to computational effects as well as effects of non-hydrostaticity in the measurements.
The fact that the calculations on the \ce{MoP2}-type model for the high-pressure phase are consistent with the results by Li et al. provides further evidence that the \ce{MoP2}-type structure is correct. The band structure of the AFM1 structure at 12~GPa is shown in Fig.~\ref{fig:bands} both with and without
Spin Orbit Coupling (SOC). 
\begin{figure*}[h]
    \centering
    \includegraphics[width=\textwidth]{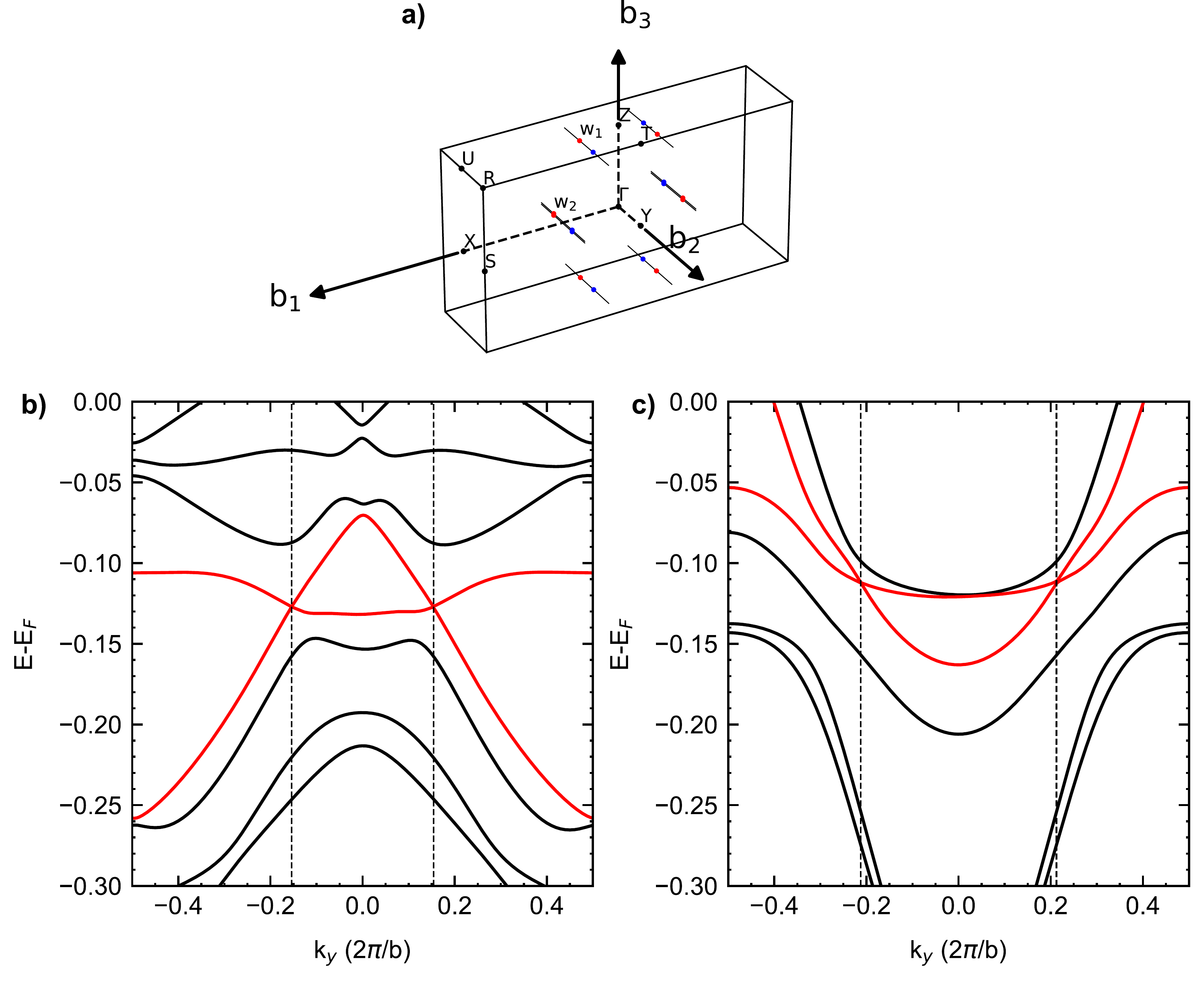}
    \caption{a) Brillouin zone of the HP-II AFM1 structure at 12 GPa with Weyl points
    $w_1$ and $w_2$. Red points have positive chirality, blue points have negative chirality.
    The $w_2$ points are very close to the b$_3$=0 ($k_z$=0) plane. b) Band structure passing through the $w_1$ points. c) Band structure passing through the $w_2$ points.}\label{fig:BZWeyl}
\end{figure*}
The effect of SOC is minor in this system and it results in narrow splittings
of the bands in the vicinity of the Fermi level. At 12~GPa the system is a semimetal with a pseudogap and low density of states at the Fermi level (Fig.~S3). The AFM1 magnetic order removes the \emph{C} centering from the $Cmc2_1$ space group. Even though the space group is non-symmorphic, we verified that the system is not an \emph{altermagnet}~\cite{Yuan2021,Mazin2021}, i.e. that no spin-splitting occurs at any k-point of the Brillouin zone. 
The large number of band crossings or near-crossings around the Fermi level, could reveal the presence of Weyl points. Similarly to the parent structure compounds, \ce{MoP2} and \ce{WP2}~\cite{MoP2WP2_weyl},
we found two symmetry independent type-I Weyl points. The results are reported in Table~\ref{tab:weylpoints}.
From Fig.~\ref{fig:BZWeyl} a) $w_1$ points are located near to the top and bottom face of the Brillouin zone. Due to the orthorhombic symmetry, $w_1$ points with opposite chirality sit aligned along the $b_2$ and $b_3$ reciprocal lattice vectors. The $w_2$ points are instead located near the $k_z=0$ ($b_3=0$) plane and are composed by very close pairs, as in \ce{MoP2}. As the $w_2$ points have the same chirality
they are robust with respect to strain and small deformations. The band structure passing through the $w_1$ and $w_2$ points reveals tilted cones (Fig.~\ref{fig:BZWeyl} b-c)), with just one point crossing the constant energy plane (type-I Weyl points).
\begin{table*}[h]
\centering
\caption{Coordinates, chirality and energy of Weyl points in the HP-II AFM1 structure at 12 GPa. For each point listed, the remaining 7 ones are generated by applying the mirror symmetries.}
\begin{tabular}{cccccc}
\toprule
Weyl point & k$_x$ (Å$^{-1}$) & k$_y$ (Å$^{-1}$) & k$_z$ (Å$^{-1}$) & Chirality & E$-$E$_F$ (eV)\\
\midrule
$w_1$ & 0.2141 & 0.0792 & 0.4625 & $+$1 & $-$0.1271 \\
$w_2$ & 0.3729 & 0.1096 & 0.0055 & $+$1 & $-$0.1129 \\
\bottomrule
\end{tabular}

\label{tab:weylpoints}
\end{table*}

\subsection{Equation of state of the high pressure polymorphs of \ce{CrSb2}}
\begin{figure*}[h!]
     \centering   \includegraphics[width=\textwidth]{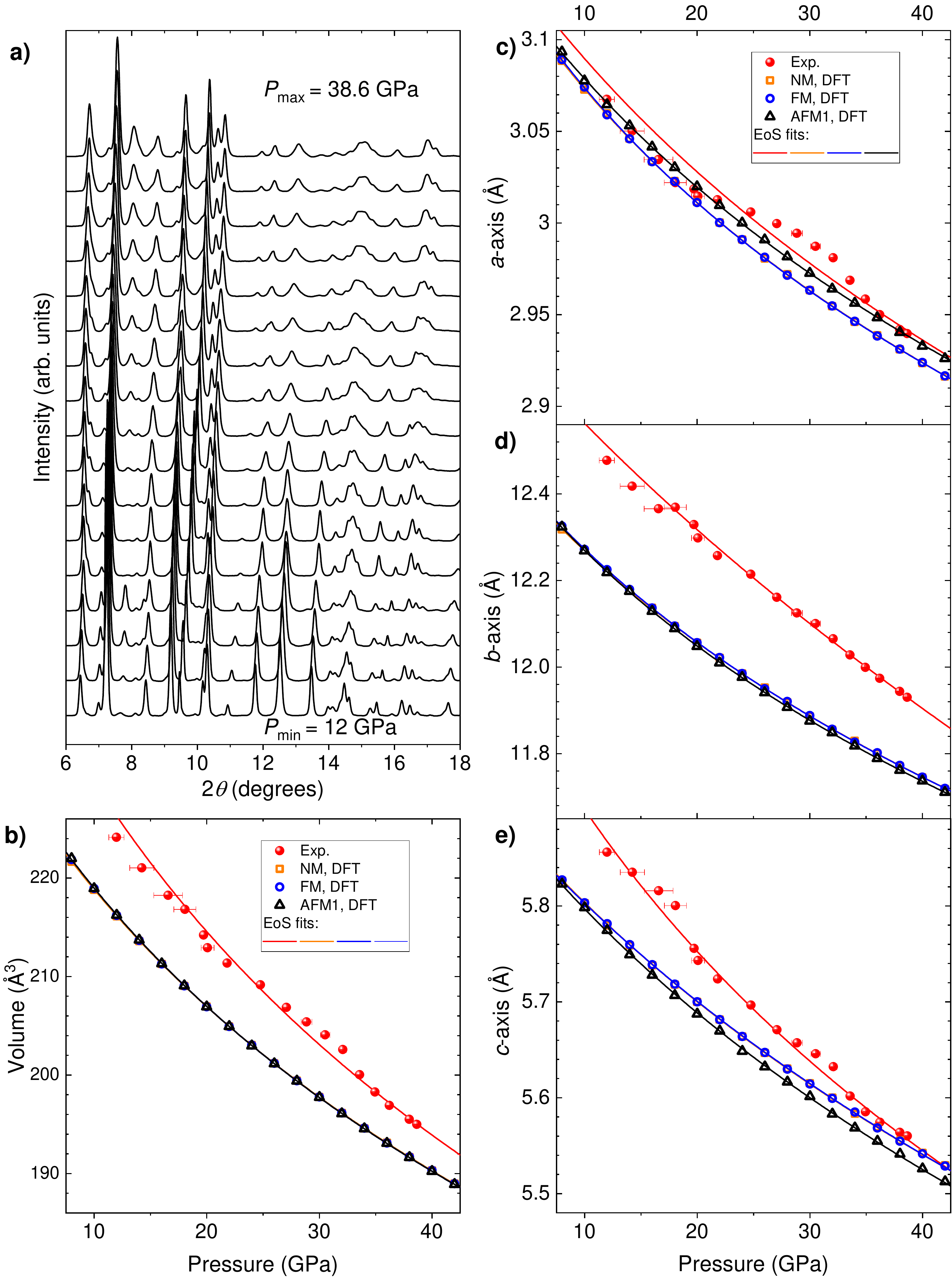}
\caption{a) Background subtracted waterfall plot of the diffractograms for the HP-II phase. b-e) Experimental and calculated unit cell parameters for the HP-II structure.}   \label{fig:MoP2_all}
\end{figure*}

\begin{figure*}[h!]
    \includegraphics[width=\textwidth]{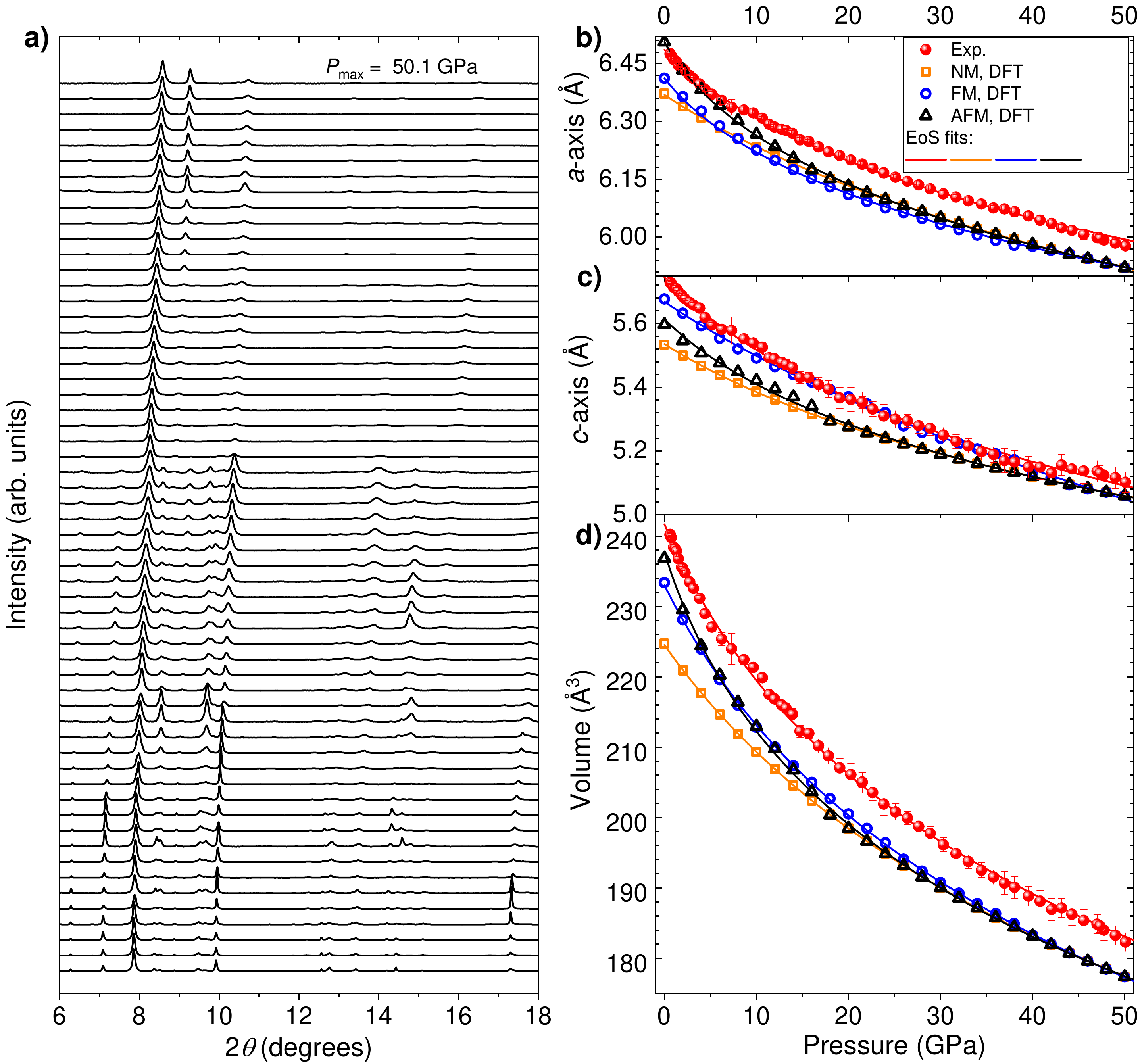}
    \caption{a) Background subtracted waterfall plot of the diffractograms for the HP-I phase. b-d) Experimental and calculated unit cell parameters for the HP-I structure}
    \label{fig:CuAl2_all}
\end{figure*}

The diffractograms for the HP-II structure are shown in Fig. \ref{fig:MoP2_all} a) and the refined unit cell parameters are shown together with the values from the DFT calculations in Fig. \ref{fig:MoP2_all} b-e) together with the Birch-Murnaghan EoS fits (Table \ref{tab:EoS} for the volume and \textcolor{red}{Table S3 in the SI for the linear moduli}).
The volume \textcolor{red}{decreased} with pressure in a scattered way, which is attributed to strain observed for the Au pressure standard as well as a change in the measured sample position at 19.3 GPa and change in the measured Au position at 21.7 GPa. From the linear moduli the $b$-axis was seen to be the least compressible while the $c$-axis was the most compressible. As the angle  between the bond of the Sb-Sb dimer and the $b$-axis and $c$-axis is approx. $45\degree$, this bond stabilizes both axes equally. Therefore, the lower compressibility of the $b$-axis is attributed to the compression of the Cr-Sb bonds, as they are oriented more along the $b$- than the $c$-axis. Along the $a$-axis both the Cr-Cr and Sb-Sb bond distances are equal to the lattice constant, and here the intermediate compressibility is attributed to the Cr-Cr interactions. The parent compounds, \ce{MoP2} and \textcolor{red}{\ce{WP2},} have a higher bulk modulus, \textcolor{red}{however the same trend for the linear moduli with} $M_{0,b}>M_{0,a}>M_{0,c}$ is also seen for \ce{MoP2} \textcolor{red}{and \ce{WP2}} \cite{WP2_high_pressure, <MoP2>}. \\
\\
For the HP-I structure the diffractograms are shown in Fig.  \ref{fig:CuAl2_all} a). The refined unit cell parameters and the unit cell parameters calculated by DFT are shown in Fig. \ref{fig:CuAl2_all}c) together with the Birch-Murnaghan EoS fits (Table \ref{tab:EoS} for the volume and \textcolor{red}{Table S4 in the SI for the linear moduli}). The $a$- and $c$-axis were seen to decrease relatively smoothly with pressure, while for the volume some points around 5-\SI{15}{GPa} were scattered around the fit.
The obtained ambient pressure unit cell values (Table S4 and Table \ref{tab:EoS} for the volume) are in good agreement with the values obtained by Takizawa et al. and Jiao et al. \cite{<Takizawa_1>,<HPCrSb2Takizawa>,<Jiao_HP_I>}.

\begin{table*}[h]
    \centering
    \caption{\label{tab:EoS}Equation of state parameters for both high pressure polymorphs of \ce{CrSb2} as well as for related structures.}
    \begin{threeparttable}
    \begin{tabular}{lllll}
    \toprule
&$V_0$ (Å$^3$) & $K_0$ (GPa)&$K'_0$&Comments\\
\midrule
\multicolumn{2}{l}{\underline{\ce{CuAl2}-type structures:}}\\
HP-I \ce{CrSb2}&241.68(15)& 77(2)& 6.2(3)& This study\\
\ce{TiSb2} && 91\tnote{a}&& Tavassoli et al. \cite{TiSb2}\\
\ce{V_{0.97}Sb2}&&85\tnote{a}&& Failamani et al. \cite{VSb2}\\
\ce{CuAl2} &179.5(6)& 117(13)&4 (fixed)& Grin et al. \cite{cuAl2}\\
\midrule
\multicolumn{2}{l}{\underline{\ce{MoP2}-type structures:}}\\
HP-II \ce{CrSb2}&251.8(14)&91(3)&4 (fixed)& This study\\
\ce{MoP2}&&238(20) & & Soto et al.           \cite{<MoP2>}\\
\textcolor{red}{\ce{WP2}}& \textcolor{red}{174.8(2)} & \textcolor{red}{234.0(34)}& \textcolor{red}{4(fixed)} & \textcolor{red}{Chi et al. \cite{WP2_high_pressure}}\\
\midrule
\multicolumn{2}{l}{\underline{Marcasite-type:}}\\
\ce{CrSb2} &137.2(2)&33(3)&13.6(19)&Ehrenreich-Petersen et al. \cite{<EmmaMoP2>}\\ &130.7(2)&74.2(2)&4 (fixed)& Li et al. \cite{KinesiskHP-II}\\    \bottomrule\end{tabular}
    \begin{tablenotes}
        \item[a]{Measured by resonant ultrasound spectroscopy} 
    \end{tablenotes}
        \end{threeparttable}
\end{table*}
From the obtained linear moduli the $c$-axis was seen to be more compressible \textcolor{red}{than the $a$-axis}, which also has been reported for \ce{TiSb2} by Armbrüster et al. which they attributed to weaker Ti-Ti bonds compared to the Sb-Sb bonds \cite{TiSb2_VSb2}.
By comparing the HP-I phase to its isostructural compounds \ce{TiSb2} and \ce{V_{1-x}Sb2} a decreasing trend in bulk modulus (Table \ref{tab:EoS}) is seen going from Ti-V-Cr. As the bulk modulus of the elements increases from Ti-V-Cr \cite{Ti_Vohra,V_EoS,Cr_EoS}, and the $M$-Sb distance ($M$=Ti, V, Cr) decreases when going from Ti-V-Cr an increasing trend in the bulk modulus would be expected. However, the Sb-Sb distance decreases when going in this direction, and as the Sb-Sb bond had been shown to be the least compressible, this increase in bond distance explains the decreasing bulk modulus and furthermore verifies that the Sb-Sb bond is the least compressible \cite{<Takizawa_1>,TiSb2_VSb2}.

\subsection{Discussion of bulk moduli among the polymorphs of \ce{CrSb2}}
In Table \ref{tab:bonddist} selected bond distances for the three polymorphs are shown. To facilitate a better comparison among the polymorphs, the atomic coordinates from the SC-XRD experiment were used for the unit cell extrapolated to 0 GPa for the HP-II phase. At 0 GPa the Cr-Sb distance in the HP-II  polymorph is similar to the Cr-Sb distance in the marcasite phase, where the coordination number is six. Thereby the higher bulk-modulus of the HP-II phase can be explained by its higher coordination number of seven. In the HP-I phase Cr is coordinated to eight Sb atoms, however the bulk modulus for this polymorph is lower than for the HP-II phase, which is attributed to the longer Cr-Sb distance in the HP-I polymorph of 2.834 Å \cite{<HPCrSb2Takizawa>}.
As the HP-II structure is not stable at 0 GPa, the  bulk moduli are also compared at 12 GPa, where the bulk modulus for the HP-II phase is 136(3) GPa and for the HP-I phase 145(1) GPa. Here, the Cr-Sb bond length is still shorter for the HP-II phase, and the slightly higher bulk modulus for the HP-I phase can be explained by its shorter Cr-Cr bond distance.

\begin{table}[h]
    \centering
    \caption{\label{tab:bonddist}Bond distances of the different polymorphs of \ce{CrSb2}}
   \begin{threeparttable}
    \begin{tabular}{lllll}
    \toprule
    &Cr-Cr (Å)&Cr-Sb (Å)& Sb-Sb (Å)&Reference\\
    \midrule
    \underline{0 GPa}\\
    \multirow{2}{*}{Marcasite-type}&3.2715(7)&2.689(9) ($\times$4) &2.84(2)&\cite{<CrSb2_unitcell>}\\
                                    &&2.746(8) ($\times$2) &&\\
    \midrule
    HP-I (\ce{CuAl2}-type)&2.8720(2)&2.8339(14)&2.914(9)&\cite{<HPCrSb2Takizawa>}\\
    \midrule
    \multirow{5}{*}{HP-II (\ce{MoP2}-type)\tnote{a}}&3.172(13)&2.714(11)&2.907(7)&This study\\
                                &           &   2.729(7) ($\times$2)     &        &\\
                                &           &   2.772(9) ($\times$2)    &&\\
                                &           &   2.743(6) ($\times$2)       &        &\\
                                &           &           &        & \\    
    \toprule
    \underline{12 GPa}\\
    HP-I (\ce{CuAl2}-type)&2.7413(17)&2.7298(7)&2.8431(7)&This study\\
    \hline
    \multirow{5}{*}{HP-II (\ce{MoP2}-type)}&3.0751(2)&2.624(7)&2.803(3)&This study\\
                        &&2.645(4) ($\times$2)&  &\\
                        &&2.635(4) ($\times$2)&&\\
                        &&
    2.683(6) ($\times$2)&&\\
        
        \bottomrule
    \end{tabular}
    \begin{tablenotes}
        \item[a]{The values at 0 GPa were obtained using the $L_0$ values from table S3 and atomic coordinates obtained through SC-XRD.} 
    \end{tablenotes}
    \end{threeparttable}
\end{table}

\section{Conclusion}
Using single crystal X-ray diffraction at 12 GPa it was shown that the high pressure phase of \ce{CrSb2} crystallizes in the orthorombic $Cmc2_1$ space group, with a \ce{MoP2}-type structure with unit cell constants of $a=3.0752(2)$ Å, $b=12.4476(19)$ Å and $c=5.8741(2)$ Å well in accordance with the obtained unit cell from DFT calculations and Le Bail refinements of powder diffractograms. Furthermore, electrical resistance measurements under pressure had shown that the phase transition is accompanied by a metallization of the system, and the experimentally measured resistance was verified by resistivity calculations. Furthermore, the calculations showed that the Hall coefficient changes from positive to negative during the phase transition from the marcasite to the \ce{MoP2}-type phase. 

The calculated band-structure at 12 GPa has shown that this polymorph is a robust type-I Weyl semimetal. Furthermore, the compressibility of both high pressure polymorphs of \ce{CrSb2} were investigated, showing no further phase transitions for both polymorphs,  and Le Bail refinements of the unit cell parameters showed that the \ce{MoP2} phase was less compressible than the \ce{CuAl2}-type phase which was attributed to the longer Cr-Sb distance in the \ce{CuAl2}-type polymorph.

\section{Acknowledgements}
This research was funded by the Independent Research Fund Denmark (7027-00077B and 1026-00409B) and we furthermore acknowledge the Danish Agency for Science, Technology, and Innovation for funding the instrument center DanScatt. We acknowledge MAX IV Laboratory for beamtime on the  DanMAX beamline (Proposal 20211063). Research conducted at MAX IV, a Swedish national user facility, is supported by the Swedish Research council under contract 2018-07152, the Swedish Governmental Agency for Innovation Systems under contract 2018-04969, and Formas under contract 2019-02496. DanMAX is funded by the NUFI grant no. 4059-00009B. Portions of this work were performed at GeoSoilEnviroCARS (The University of Chicago, Sector 13), Advanced Photon Source (APS), Argonne National Laboratory. GeoSoilEnviroCARS is supported by the National Science Foundation – Earth Sciences (EAR – 1634415). This research used resources of the Advanced Photon Source, a U.S. Department of Energy (DOE) Office of Science User Facility operated for the DOE Office of Science by Argonne National Laboratory under Contract No. DE-AC02-06CH11357. Use of the COMPRES-GSECARS gas loading system was supported by COMPRES under NSF Cooperative Agreement EAR -1606856 and by GSECARS through NSF grant EAR-1634415 and DOE grant DE-FG02-94ER14466. This research used resources of the Advanced Photon Source, a U.S. Department of Energy (DOE) Office of Science User Facility operated for the DOE Office of Science by Argonne National Laboratory under Contract No. DE-AC02-06CH11357. We acknowledge DESY (Hamburg, Germany), a member of the Helmholtz Association HGF, for the provision of experimental facilities. Parts of this research were carried out at PETRA III using the P02.2 beamline. Beamtime was allocated for proposal H-20010016. DC acknowledges support from ICSC -- Centro Nazionale di Ricerca in High Performance Computing, Big Data and Quantum Computing, funded by European Union -- NextGenerationEU (Grant number CN00000013).

\bibliography{references}

\begin{thebibliography}{10}
\expandafter\ifx\csname url\endcsname\relax
  \def\url#1{\texttt{#1}}\fi
\expandafter\ifx\csname urlprefix\endcsname\relax\def\urlprefix{URL }\fi
\expandafter\ifx\csname href\endcsname\relax
  \def\href#1#2{#2} \def\path#1{#1}\fi

\bibitem{alphawP2}
F.~Hulliger, New representatives of the \ce{NbAs2} and \ce{ZrAs2} structures,
  Nature 204 (Nov 1964).
\newblock \href {https://doi.org/10.1038/204775a0}
  {\path{doi:10.1038/204775a0}}.

\bibitem{MoP2WP2}
S.~Rundqvist, T.~Lundström, X-ray studies of molybdenum and tungsten
  phosphides, Acta Chemica Scandinavica 17 (1963) 37--46.
\newblock \href {https://doi.org/10.3891/acta.chem.scand.17-0037}
  {\path{doi:10.3891/acta.chem.scand.17-0037}}.

\bibitem{MoP2WP2_weyl}
G.~Aut\`es, D.~Gresch, M.~Troyer, A.~A. Soluyanov, O.~V. Yazyev, Robust
  type-{II} weyl semimetal phase in transition metal diphosphides ${X}$\ce{P2}
  (${X}=${Mo, W}), Physical Review Letters 117 (2016) 066402.
\newblock \href {https://doi.org/10.1103/PhysRevLett.117.066402}
  {\path{doi:10.1103/PhysRevLett.117.066402}}.

\bibitem{alphawp2MR}
J.~Du, Z.~Lou, S.~Zhang, Y.~Zhou, B.~Xu, Q.~Chen, Y.~Tang, S.~Chen, H.~Chen,
  Q.~Zhu, H.~Wang, J.~Yang, Q.~Wu, O.~V. Yazyev, M.~Fang, Extremely large
  magnetoresistance in the topologically trivial semimetal
  $\ensuremath{\alpha}\ensuremath{-}$\ce{WP2}, Physical Review B 97 (2018)
  245101.
\newblock \href {https://doi.org/10.1103/PhysRevB.97.245101}
  {\path{doi:10.1103/PhysRevB.97.245101}}.

\bibitem{KinesiskHP-II}
C.~Li, K.~Liu, S.~Peng, Q.~Feng, D.~Jiang, T.~Wen, H.~Xiao, B.~Yue, Y.~Wang,
  Rewritable pressure-driven n–p conduction switching in marcasite-type
  \protect{CrSb$_2$}, Chemistry of Materials 35~(3) (2023) 1449--1457.
\newblock \href {https://doi.org/10.1021/acs.chemmater.2c03673}
  {\path{doi:10.1021/acs.chemmater.2c03673}}.

\bibitem{<EmmaMoP2>}
E.~Ehrenreich-Petersen, M.~F. Hansen, J.~Jeanneau, D.~Ceresoli, F.~Menescardi,
  M.~Ottesen, V.~Prakapenka, S.~N. Tkachev, M.~Bremholm, Seven-coordinated
  high-pressure phase of \ce{CrSb2} and experimental equation of state of
  \ce{MSb2} (\ce{M} =\ce{Cr}, \ce{Fe}, \ce{Ru}, \ce{Os}), Inorganic Chemistry
  62~(31) (2023) 12203--12212.
\newblock \href {https://doi.org/10.1021/acs.inorgchem.3c00227}
  {\path{doi:10.1021/acs.inorgchem.3c00227}}.

\bibitem{<CrSb2_neutron>}
H.~Holseth, A.~Kjekshus, A.~F. Andresen, Compounds with the marcasite type
  crystal structure. vi. neutron diffraction studies of \ce{CrSb2} and
  \ce{FeSb2}, Acta Chemica Scandinavica 24 (1970) 3309--3316.
\newblock \href {https://doi.org/10.3891/acta.chem.scand.24-3309}
  {\path{doi:10.3891/acta.chem.scand.24-3309}}.

\bibitem{<Physb2012>}
B.~C. Sales, A.~F. May, M.~A. McGuire, M.~B. Stone, D.~J. Singh, D.~Mandrus,
  Transport, thermal, and magnetic properties of the narrow-gap semiconductor
  \ce{CrSb2}, Physical Review B 86 (2012) 235136.
\newblock \href {https://doi.org/10.1103/PhysRevB.86.235136}
  {\path{doi:10.1103/PhysRevB.86.235136}}.

\bibitem{CrSb2_surf_MR}
K.~Nakagawa, M.~Kimata, T.~Yokouchi, Y.~Shiomi,
  \href{https://link.aps.org/doi/10.1103/PhysRevB.107.L180405}{Surface
  anisotropic magnetoresistance in the antiferromagnetic semiconductor
  \ce{CrSb2}}, Phys. Rev. B 107 (2023) L180405.
\newblock \href {https://doi.org/10.1103/PhysRevB.107.L180405}
  {\path{doi:10.1103/PhysRevB.107.L180405}}.
\newline\urlprefix\url{https://link.aps.org/doi/10.1103/PhysRevB.107.L180405}

\bibitem{<Takizawa_1>}
H.~Takizawa, K.~Uheda, T.~Endo, M.~Shimada, High pressure phase transition of
  \ce{CrSb2}, Review of High Pressure Science and Technology/Koatsuryoku No
  Kagaku To Gijutsu 7 (1998) 1043--1045.
\newblock \href {https://doi.org/10.4131/jshpreview.7.1043}
  {\path{doi:10.4131/jshpreview.7.1043}}.

\bibitem{TiSb2_VSb2}
M.~Armbrüster, W.~Schnelle, U.~Schwarz, Y.~Grin, Chemical bonding in
  \ce{TiSb2} and \ce{VSb2}: A quantum chemical and experimental study,
  Inorganic Chemistry 46~(16) (2007) 6319--6328.
\newblock \href {https://doi.org/10.1021/ic070284p}
  {\path{doi:10.1021/ic070284p}}.

\bibitem{TiSb2_dirac}
W.~Xia, X.~Shi, Y.~Wang, W.~Ge, H.~Su, Q.~Wang, X.~Wang, N.~Yu, Z.~Zou, Y.~Hao,
  W.~Zhao, Y.~Guo, \href{https://doi.org/10.1063/5.0001566}{{The de Haas-van
  Alphen quantum oscillations in a three-dimensional Dirac semimetal
  \ce{TiSb2}}}, Applied Physics Letters 116~(14) (2020) 142103.
\newblock \href {https://doi.org/10.1063/5.0001566}
  {\path{doi:10.1063/5.0001566}}.
\newline\urlprefix\url{https://doi.org/10.1063/5.0001566}

\bibitem{<HPCrSb2Takizawa>}
H.~Takizawa, K.~Uheda, T.~Endo, A new ferromagnetic polymorph of \ce{CrSb2}
  synthesized under high pressure, Journal of Alloys and Compounds 287~(1)
  (1999) 145--149.
\newblock \href {https://doi.org/https://doi.org/10.1016/S0925-8388(99)00056-0}
  {\path{doi:https://doi.org/10.1016/S0925-8388(99)00056-0}}.

\bibitem{<Jiao_HP_I>}
Y.~Y. Jiao, Z.~Y. Liu, M.~A. McGuire, S.~Calder, J.-Q. Yan, B.~C. Sales, J.~P.
  Sun, Q.~Cui, N.~N. Wang, Y.~Sui, Y.~Uwatoko, B.~S. Wang, X.~L. Dong, J.-G.
  Cheng, High-pressure phase of \ce{CrSb2}: A new quasi-one-dimensional
  itinerant magnet with competing interactions, Physical Review Materials 3
  (2019) 074404.
\newblock \href {https://doi.org/10.1103/PhysRevMaterials.3.074404}
  {\path{doi:10.1103/PhysRevMaterials.3.074404}}.

\bibitem{Mr-wP2_MoP2}
N.~Kumar, Y.~Sun, N.~Xu, K.~Manna, M.~Yao, V.~Süss, I.~Leermakers, O.~Young,
  T.~Förster, M.~Schmidt, H.~Borrmann, B.~Yan, U.~Zeitler, M.~Shi, C.~Felser,
  C.~Shekhar, Extremely high magnetoresistance and conductivity in the
  type-{II} weyl semimetals \ce{WP2} and \ce{MoP2}, Nature Communications 8
  (Nov 2017).
\newblock \href {https://doi.org/10.1038/s41467-017-01758-z}
  {\path{doi:10.1038/s41467-017-01758-z}}.

\bibitem{WP2_HHG}
Y.-Y. Lv, J.~Xu, S.~Han, C.~Zhang, Y.~Han, J.~Zhou, S.-H. Yao, X.-P. Liu, M.-H.
  Lu, H.~Weng, Z.~Xie, Y.~B. Chen, J.~Hu, Y.-F. Chen, S.~Zhu, High-harmonic
  generation in weyl semimetal $\beta$-\ce{WP2} crystals, Nature Communications
  12 (2021) 6437.
\newblock \href {https://doi.org/https://doi.org/10.1038/s41467-021-26766-y}
  {\path{doi:https://doi.org/10.1038/s41467-021-26766-y}}.

\bibitem{Amplifier}
A.~Toniato, B.~Gotsmann, E.~Lind, C.~B. Zota, Weyl semi-metal-based
  high-frequency amplifiers, in: 2019 IEEE International Electron Devices
  Meeting (IEDM), 2019, pp. 9.4.1--9.4.4.
\newblock \href {https://doi.org/10.1109/IEDM19573.2019.8993575}
  {\path{doi:10.1109/IEDM19573.2019.8993575}}.

\bibitem{Alpha_MoP2}
X.~Liu, Z.~Yu, J.~Li, Z.~Xu, C.~Zhou, Z.~Dong, L.~Zhang, X.~Wang, N.~Yu,
  Z.~Zou, X.~Wang, Y.~Guo, \href{https://dx.doi.org/10.1088/1674-1056/ac633d}{A
  new transition metal diphosphide $\alpha$-\ce{MoP2} synthesized by a
  high-temperature and high-pressure technique}, Chinese Physics B 32~(1)
  (2023) 018102.
\newblock \href {https://doi.org/10.1088/1674-1056/ac633d}
  {\path{doi:10.1088/1674-1056/ac633d}}.
\newline\urlprefix\url{https://dx.doi.org/10.1088/1674-1056/ac633d}

\bibitem{WP2_high_pressure}
Z.~Chi, J.~Zhang, Z.~Gong, F.~Peng, X.~Wang, G.~Dong, Y.~Li, Y.~Shi, Y.~Ge,
  X.~Yang, Z.~Zhang, G.~Xu, N.~Hao, C.~Zhou, J.~Qin,
  \href{https://www.sciencedirect.com/science/article/pii/S2542529324000488}{{Pressure-induced
  Lifshitz transition in the type-II Weyl semimetal \ce{WP2}}}, Materials Today
  Physics 42 (2024) 101372.
\newblock \href {https://doi.org/https://doi.org/10.1016/j.mtphys.2024.101372}
  {\path{doi:https://doi.org/10.1016/j.mtphys.2024.101372}}.
\newline\urlprefix\url{https://www.sciencedirect.com/science/article/pii/S2542529324000488}

\bibitem{COMPRES}
K.~D. Leinenweber, J.~A. Tyburczy, T.~G. Sharp, E.~Soignard, T.~Diedrich, W.~B.
  Petuskey, Y.~Wang, J.~L. Mosenfelder, Cell assemblies for reproducible
  multi-anvil experiments (the {COMPRES} assemblies), American Mineralogist
  97~(2-3) (2012) 353--368.
\newblock \href {https://doi.org/10.2138/am.2012.3844}
  {\path{doi:10.2138/am.2012.3844}}.

\bibitem{P022}
H.-P. Liermann, Z.~Kon{\^{o}}pkov{\'{a}}, W.~Morgenroth, K.~Glazyrin,
  J.~Bednar{\v{c}}ik, E.~E. McBride, S.~Petitgirard, J.~T. Delitz, M.~Wendt,
  Y.~Bican, A.~Ehnes, I.~Schwark, A.~Rothkirch, M.~Tischer, J.~Heuer,
  H.~Schulte-Schrepping, T.~Kracht, H.~Franz, {The Extreme Conditions Beamline
  P02.2 and the~Extreme Conditions Science Infrastructure at~PETRAIII}, Journal
  of Synchrotron Radiation 22~(4) (2015) 908--924.
\newblock \href {https://doi.org/10.1107/S1600577515005937}
  {\path{doi:10.1107/S1600577515005937}}.

\bibitem{<Syassen>}
K.~Syassen, Ruby under pressure, High Pressure Research 28~(2) (2008) 75--126.
\newblock \href {https://doi.org/10.1080/08957950802235640}
  {\path{doi:10.1080/08957950802235640}}.

\bibitem{cryspro}
CrysAlisPRO, Oxford Diffraction /Agilent Technologies UK Ltd, Yarnton, England.

\bibitem{Olex2}
O.~V. Dolomanov, L.~J. Bourhis, R.~J. Gildea, J.~A.~K. Howard, H.~Puschmann,
  {{\it OLEX2}: a complete structure solution, refinement and analysis
  program}, Journal of Applied Crystallography 42~(2) (2009) 339--341.
\newblock \href {https://doi.org/10.1107/S0021889808042726}
  {\path{doi:10.1107/S0021889808042726}}.

\bibitem{Shelxl}
G.~M. Sheldrick, {Crystal structure refinement with {\it SHELXL}}, Acta
  Crystallographica Section C 71~(1) (2015) 3--8.
\newblock \href {https://doi.org/10.1107/S2053229614024218}
  {\path{doi:10.1107/S2053229614024218}}.

\bibitem{Shelxt}
G.~M. Sheldrick, {{\it SHELXT} {--} Integrated space-group and
  crystal-structure determination}, Acta Crystallographica Section A 71~(1)
  (2015) 3--8.
\newblock \href {https://doi.org/10.1107/S2053273314026370}
  {\path{doi:10.1107/S2053273314026370}}.

\bibitem{<GasLoad>}
M.~Rivers, V.~B. Prakapenka, A.~Kubo, C.~Pullins, C.~M. Holl, S.~D. Jacobsen,
  The \protect{COMPRES/GSECARS} gas-loading system for diamond anvil cells at
  the advanced photon source, High Pressure Research 28~(3) (2008) 273--292.
\newblock \href {https://doi.org/10.1080/08957950802333593}
  {\path{doi:10.1080/08957950802333593}}.

\bibitem{<CuEoS>}
A.~Dewaele, P.~Loubeyre, M.~Mezouar, Equations of state of six metals above 94
  {GPa}, Physical Review B 70 (2004) 094112.
\newblock \href {https://doi.org/10.1103/PhysRevB.70.094112}
  {\path{doi:10.1103/PhysRevB.70.094112}}.

\bibitem{AuEoS}
Y.~Fei, A.~Ricolleau, M.~Frank, K.~Mibe, G.~Shen, V.~Prakapenka, Toward an
  internally consistent pressure scale, Proceedings of the National Academy of
  Sciences 104~(22) (2007) 9182--9186.
\newblock \href {https://doi.org/10.1073/pnas.0609013104}
  {\path{doi:10.1073/pnas.0609013104}}.

\bibitem{<Dioptas>}
C.~Prescher, V.~B. Prakapenka, \protect{DIOPTAS}: a program for reduction of
  two-dimensional x-ray diffraction data and data exploration, High Pressure
  Research 35~(3) (2015) 223--230.
\newblock \href {https://doi.org/10.1080/08957959.2015.1059835}
  {\path{doi:10.1080/08957959.2015.1059835}}.

\bibitem{Fullprof}
J.~Rodríguez-Carvajal, Recent advances in magnetic structure determination by
  neutron powder diffraction, Physica B: Condensed Matter 192~(1) (1993)
  55--69.
\newblock \href {https://doi.org/https://doi.org/10.1016/0921-4526(93)90108-I}
  {\path{doi:https://doi.org/10.1016/0921-4526(93)90108-I}}.

\bibitem{EoSfitGUI}
J.~Gonzalez-Platas, M.~Alvaro, F.~Nestola, R.~Angel, {{\it EosFit7-GUI}: a new
  graphical user interface for equation of state calculations, analyses and
  teaching}, Journal of Applied Crystallography 49~(4) (2016) 1377--1382.
\newblock \href {https://doi.org/10.1107/S1600576716008050}
  {\path{doi:10.1107/S1600576716008050}}.

\bibitem{<vesta>}
K.~Momma, F.~Izumi, {{\it VESTA3} for three-dimensional visualization of
  crystal, volumetric and morphology data}, Journal of Applied Crystallography
  44~(6) (2011) 1272--1276.
\newblock \href {https://doi.org/10.1107/S0021889811038970}
  {\path{doi:10.1107/S0021889811038970}}.

\bibitem{<First_designeranvil>}
S.~T. Weir, J.~Akella, C.~Aracne-Ruddle, Y.~K. Vohra, S.~A. Catledge, Epitaxial
  diamond encapsulation of metal microprobes for high pressure experiments,
  Applied Physics Letters 77~(21) (2000) 3400--3402.
\newblock \href {https://doi.org/10.1063/1.1326838}
  {\path{doi:10.1063/1.1326838}}.

\bibitem{<quantespresso2009>}
P.~Giannozzi, S.~Baroni, N.~Bonini, M.~Calandra, R.~Car, C.~Cavazzoni,
  D.~Ceresoli, G.~L. Chiarotti, M.~Cococcioni, I.~Dabo, A.~D. Corso,
  S.~de~Gironcoli, S.~Fabris, G.~Fratesi, R.~Gebauer, U.~Gerstmann,
  C.~Gougoussis, A.~Kokalj, M.~Lazzeri, L.~Martin-Samos, N.~Marzari, F.~Mauri,
  R.~Mazzarello, S.~Paolini, A.~Pasquarello, L.~Paulatto, C.~Sbraccia,
  S.~Scandolo, G.~Sclauzero, A.~P. Seitsonen, A.~Smogunov, P.~Umari, R.~M.
  Wentzcovitch, {QUANTUM} {ESPRESSO}: a modular and open-source software
  project for quantum simulations of materials, Journal of Physics: Condensed
  Matter 21~(39) (2009) 395502.
\newblock \href {https://doi.org/10.1088/0953-8984/21/39/395502}
  {\path{doi:10.1088/0953-8984/21/39/395502}}.

\bibitem{<quantespresso2017>}
P.~Giannozzi, O.~Andreussi, T.~Brumme, O.~Bunau, M.~B. Nardelli, M.~Calandra,
  R.~Car, C.~Cavazzoni, D.~Ceresoli, M.~Cococcioni, N.~Colonna, I.~Carnimeo,
  A.~D. Corso, S.~de~Gironcoli, P.~Delugas, R.~A. DiStasio, A.~Ferretti,
  A.~Floris, G.~Fratesi, G.~Fugallo, R.~Gebauer, U.~Gerstmann, F.~Giustino,
  T.~Gorni, J.~Jia, M.~Kawamura, H.-Y. Ko, A.~Kokalj, E.~Kü{\c{c}}ükbenli,
  M.~Lazzeri, M.~Marsili, N.~Marzari, F.~Mauri, N.~L. Nguyen, H.-V. Nguyen,
  A.~O. de-la Roza, L.~Paulatto, S.~Ponc{\'{e}}, D.~Rocca, R.~Sabatini,
  B.~Santra, M.~Schlipf, A.~P. Seitsonen, A.~Smogunov, I.~Timrov,
  T.~Thonhauser, P.~Umari, N.~Vast, X.~Wu, S.~Baroni, Advanced capabilities for
  materials modelling with {QUANTUM} {ESPRESSO}, Journal of Physics: Condensed
  Matter 29~(46) (2017) 465901.
\newblock \href {https://doi.org/10.1088/1361-648x/aa8f79}
  {\path{doi:10.1088/1361-648x/aa8f79}}.

\bibitem{<functionale>}
J.~P. Perdew, A.~Ruzsinszky, G.~I. Csonka, O.~A. Vydrov, G.~E. Scuseria, L.~A.
  Constantin, X.~Zhou, K.~Burke, Restoring the density-gradient expansion for
  exchange in solids and surfaces, Physical Review Letters 100 (2008) 136406.
\newblock \href {https://doi.org/10.1103/PhysRevLett.100.136406}
  {\path{doi:10.1103/PhysRevLett.100.136406}}.

\bibitem{Kuhn2013}
G.~Kuhn, S.~Mankovsky, H.~Ebert, M.~Regus, W.~Bensch, Electronic structure and
  magnetic properties of \protect{CrSb$_2$} and \protect{FeSb$_2$} investigated
  via ab initio calculations, Physical Review B 87~(8) (feb 2013).
\newblock \href {https://doi.org/10.1103/physrevb.87.085113}
  {\path{doi:10.1103/physrevb.87.085113}}.

\bibitem{boltztrap2}
G.~K. Madsen, J.~Carrete, M.~J. Verstraete, {BoltzTraP}2, a program for
  interpolating band structures and calculating semi-classical transport
  coefficients, Computer Physics Communications 231 (2018) 140--145.
\newblock \href {https://doi.org/10.1016/j.cpc.2018.05.010}
  {\path{doi:10.1016/j.cpc.2018.05.010}}.

\bibitem{PAOFLOW2}
F.~T. Cerasoli, A.~R. Supka, A.~Jayaraj, M.~Costa, I.~Siloi,
  J.~S{\l}awi{\'{n}}ska, S.~Curtarolo, M.~Fornari, D.~Ceresoli, M.~B. Nardelli,
  Advanced modeling of materials with {PAOFLOW} 2.0: New features and software
  design, Computational Materials Science 200 (2021) 110828.
\newblock \href {https://doi.org/10.1016/j.commatsci.2021.110828}
  {\path{doi:10.1016/j.commatsci.2021.110828}}.

\bibitem{Z2PACK}
D.~Gresch, G.~Aut{\`{e}}s, O.~V. Yazyev, M.~Troyer, D.~Vanderbilt, B.~A.
  Bernevig, A.~A. Soluyanov, Z2pack: Numerical implementation of hybrid wannier
  centers for identifying topological materials, Physical Review B 95~(7) (feb
  2017).
\newblock \href {https://doi.org/10.1103/physrevb.95.075146}
  {\path{doi:10.1103/physrevb.95.075146}}.

\bibitem{<FeSb2Eg>}
H.~R. Aliabad, S.~Rabbanifar, M.~Khalid, Structural, optoelectronic and
  thermoelectric properties of \ce{FeSb2} under pressure: Bulk and monolayer,
  Physica B: Condensed Matter 570 (2019) 100--109.
\newblock \href {https://doi.org/https://doi.org/10.1016/j.physb.2019.06.001}
  {\path{doi:https://doi.org/10.1016/j.physb.2019.06.001}}.

\bibitem{<FeSb2egexp>}
A.~Mani, J.~Janaki, A.~T. Satya, T.~G. Kumary, A.~Bharathi, The pressure
  induced insulator to metal transition in \ce{FeSb2}, Journal of Physics:
  Condensed Matter 24~(7) (2012) 075601.
\newblock \href {https://doi.org/10.1088/0953-8984/24/7/075601}
  {\path{doi:10.1088/0953-8984/24/7/075601}}.

\bibitem{Yuan2021}
L.-D. Yuan, Z.~Wang, J.-W. Luo, A.~Zunger, Prediction of low-\protect{Z}
  collinear and noncollinear antiferromagnetic compounds having
  momentum-dependent spin splitting even without spin-orbit coupling, Physical
  Review Materials 5~(1) (jan 2021).
\newblock \href {https://doi.org/10.1103/physrevmaterials.5.014409}
  {\path{doi:10.1103/physrevmaterials.5.014409}}.

\bibitem{Mazin2021}
I.~I. Mazin, K.~Koepernik, M.~D. Johannes, R.~Gonz{\'{a}}lez-Hern{\'{a}}ndez,
  L.~{\v{S}}mejkal, Prediction of unconventional magnetism in doped \ce{FeSb2},
  Proceedings of the National Academy of Sciences 118~(42) (oct 2021).
\newblock \href {https://doi.org/10.1073/pnas.2108924118}
  {\path{doi:10.1073/pnas.2108924118}}.

\bibitem{<MoP2>}
V.~Soto, K.~Knorr, L.~Ehm, C.~Baehtz, B.~Winkler, M.~Avalos-Borja,
  High-pressure and high-temperature powder diffraction on molybdenum
  diphosphide, \ce{MoP2}, Zeitschrift für Kristallographie - Crystalline
  Materials 219~(6) (2004) 309--313.
\newblock \href {https://doi.org/doi:10.1524/zkri.219.6.309.34642}
  {\path{doi:doi:10.1524/zkri.219.6.309.34642}}.

\bibitem{TiSb2}
A.~Tavassoli, A.~Grytsiv, F.~Failamani, G.~Rogl, S.~Puchegger, H.~Müller,
  P.~Broz, F.~Zelenka, D.~Macciò, A.~Saccone, G.~Giester, E.~Bauer,
  M.~Zehetbauer, P.~Rogl, Constitution of the binary m-sb systems (m = \ce{Ti,
  Zr, Hf}) and physical properties of \ce{MSb2}, Intermetallics 94 (2018)
  119--132.
\newblock \href
  {https://doi.org/https://doi.org/10.1016/j.intermet.2017.12.014}
  {\path{doi:https://doi.org/10.1016/j.intermet.2017.12.014}}.

\bibitem{VSb2}
F.~Failamani, P.~Broz, D.~Macciò, S.~Puchegger, H.~Müller, L.~Salamakha,
  H.~Michor, A.~Grytsiv, A.~Saccone, E.~Bauer, G.~Giester, P.~Rogl,
  Constitution of the systems \{{V,Nb,Ta}\}-{Sb} and physical properties of
  di-antimonides \{{V,Nb,Ta}\}\ce{Sb2}, Intermetallics 65 (2015) 94--110.
\newblock \href
  {https://doi.org/https://doi.org/10.1016/j.intermet.2015.05.006}
  {\path{doi:https://doi.org/10.1016/j.intermet.2015.05.006}}.

\bibitem{cuAl2}
Y.~Grin, F.~R. Wagner, M.~Armbrüster, M.~Kohout, A.~Leithe-Jasper, U.~Schwarz,
  U.~Wedig, H.~{Georg von Schnering}, \ce{CuAl2} revisited: Composition,
  crystal structure, chemical bonding, compressibility and raman spectroscopy,
  Journal of Solid State Chemistry 179~(6) (2006) 1707--1719.
\newblock \href {https://doi.org/https://doi.org/10.1016/j.jssc.2006.03.006}
  {\path{doi:https://doi.org/10.1016/j.jssc.2006.03.006}}.

\bibitem{Ti_Vohra}
Y.~K. Vohra, P.~T. Spencer, Novel $\ensuremath{\gamma}$-phase of titanium metal
  at megabar pressures, Physical Review Letters 86 (2001) 3068--3071.
\newblock \href {https://doi.org/10.1103/PhysRevLett.86.3068}
  {\path{doi:10.1103/PhysRevLett.86.3068}}.

\bibitem{V_EoS}
W.~A. Crichton, J.~Guignard, E.~Bailey, D.~P. Dobson, S.~A. Hunt, A.~R.
  Thomson, High-temperature equation of state of vanadium, High Pressure
  Research 36~(1) (2016) 16--22.
\newblock \href {https://doi.org/10.1080/08957959.2015.1123256}
  {\path{doi:10.1080/08957959.2015.1123256}}.

\bibitem{Cr_EoS}
A.~Marizy, G.~Geneste, P.~Loubeyre, B.~Guigue, G.~Garbarino, Synthesis of bulk
  chromium hydrides under pressure of up to 120 \protect{GPa}, Physical Review
  B 97 (2018) 184103.
\newblock \href {https://doi.org/10.1103/PhysRevB.97.184103}
  {\path{doi:10.1103/PhysRevB.97.184103}}.

\bibitem{<CrSb2_unitcell>}
H.~Holseth, A.~Kjekshus, Compounds with the marcasite type crystal structure.
  {II}. on the crystal structures of the binary pnictides, Acta Chemica
  Scandinavica 22 (1968) 3284--3292.
\newblock \href {https://doi.org/10.3891/acta.chem.scand.22-3284}
  {\path{doi:10.3891/acta.chem.scand.22-3284}}.

\end{thebibliography}
\end{document}